\documentclass{aastex}              
\usepackage{emulateapj5}		
\usepackage{apjfonts}			
\usepackage{epsf}
\usepackage{natbib}

\newcommand{\hi}{{H$\,$\footnotesize I}}
\newcommand{\sm}{$\sim\,$}

\newcommand{\kms}{\,km\,s$^{-1}$}
\newcommand{\df}{\mbox{DEF}}

\newcommand{\sci}{Science}     

\slugcomment{Accepted for publication by The Astrophysical Journal.}

\shorttitle{\hi\ Deficiency on the Outskirts of Virgo}
\shortauthors{Sanchis et al.}

\begin{document}
\doublespace

\title{Are the \hi\ Deficient Galaxies on the Outskirts of Virgo Recent
Arrivals?}


\author{Teresa Sanchis\altaffilmark{1}, Jos\'e M.\ Solanes\altaffilmark{2}, Eduard Salvador-Sol\'e\altaffilmark{1,3}, Pascal Fouqu\'e\altaffilmark{4,5}, and Alberto Manrique\altaffilmark{1}}
\altaffiltext{1}{Departament d'Astronomia i
Meteorologia, Universitat de Barcelona. Av.\ Diagonal, 647; 08028~Barcelona, Spain}
\altaffiltext{2}{Departament d'Enginyeria
Inform\`atica i Ma\-te\-m\`a\-ti\-ques, Universitat Rovira i Virgili. Av.\ Pa\"\i sos Catalans, 26; 43007~Ta\-rra\-go\-na, Spain}
\altaffiltext{3}{CER d'Astrof\'\i sica, F\'\i sica de Part\'\i cules i Cosmologia, Universitat de Barcelona. Av.\ Diagonal, 647;
08028~Barcelona, Spain} 
\altaffiltext{4}{European Southern Observatory, Casilla 19001, Santiago 19, Chile}
\altaffiltext{5}{Observatoire de Paris-Meudon DESPA, F--92195 Meudon CEDEX, France} 
\email{tsanchis@am.ub.es, jsolanes@etse.urv.es, eduard@am.ub.es, pfouque@eso.org, Alberto.Manrique@am.ub.es}

             
\begin{abstract} 

The presence on the Virgo cluster outskirts of spiral galaxies with gas
deficiencies as strong as those of the inner galaxies stripped by the
intracluster medium has led us to explore the possibility that some of
these peripheral objects are not newcomers. A dynamical model for the
collapse and rebound of spherical shells under the point mass and
radial flow approximations has been developed to account for the
amplitude of the motions in the Virgo I cluster (VIC) region. According
to our analysis, it is not unfeasible that galaxies far from the
cluster, including those in a gas-deficient group well to its
background, went through its core a few Gyr ago. The implications would
be: (1) that the majority of the \hi-deficient spirals in the VIC
region might have been deprived of their neutral hydrogen by
interactions with the hot intracluster medium; and (2) that objects
spending a long time outside the cluster cores might keep the gas
deficient status without altering their morphology.

\end{abstract}


\section{An Intriguing Possibility Worth Exploring}\label{introduction}

A recent characterization of the large-scale 3D distribution of the
neutral gas (\hi) deficiency around the Virgo I Cluster (VIC) region by
\citet*[hereafter Paper~I]{Sol02} has shown that there are a
significant number of galaxies with a dearth of atomic hydrogen at
large Virgocentric distances. These peripheral gas-deficient objects,
which can be observed both in the cluster front and in a probable
background group well behind the cluster core, show gaseous
deficiencies comparable in strength to those measured in the centers of
Virgo and other rich galaxy clusters.

One of the mechanisms that can most naturally account for the observed
reduction in the interstellar gas content of cluster galaxies is the
ram pressure ablation caused by the rapid motion of galaxies through
the dense intracluster medium (ICM). There is now compelling evidence
for the decisive participation of this process in the gaseous
deficiencies of spirals located in the centers of rich clusters, either
directly from observations \citep*{GH85,GJ87,DG91} ---including the
discovery of shrunken gaseous disks \citep{Cay94,Bra00} and the finding
that \hi\ deficient spirals are on very eccentric orbits
\citep{Sol01}--- or from theoretical studies that have checked the
efficiency of this mechanism \citep*[][to name only a
few]{SAP99,QMB00,Vol01}.

In the outer cluster regions the low density of the intergalactic
medium calls, in principle, for alternative gas removal mechanisms,
such as gravitational tidal interactions. Indeed, observational
evidence suggests that processes of this kind might have played an
important role on the evolution of the galactic population in distant
clusters \citep*[e.g.,][]{vDok99} due to favorable conditions for
frequent low-relative velocity encounters among the galaxies in early
epochs. Although in the VIC region low-relative-velocity galaxy-galaxy
interactions may also be responsible for the gaseous deficiencies
observed in some of the peripheral galaxies, it should not be forgotten
that the dynamics of the Virgo region is dominated by large-scale
non-Hubble radial streaming motions. In this context, it is plausible
that some galaxies at large Virgocentric distances are on very
eccentric orbits that carry them right through the cluster center with
high relative velocities and are therefore liable to have suffered a
strong interaction with the ICM.

One of the most influential studies of the Local Supercluster based on
dynamical model calculations is the analysis by \citet{TS84} of the
infall of galaxies in the Virgo Southern Extension (or Virgo II cloud)
toward the VIC. The lumpy distribution of galaxies in space led these
authors to predict a very irregular infall rate which would be
responsible for the secular evolution of the mix of morphological types
in the cluster. It was suggested that the formation of the cluster took
place at an early epoch ---when the universe was about one fourth of
its present age \citetext{R.~B.\ Tully 2002, private communication}---
by a first generation of, probably, early-type galaxies. Afterwards,
infall was reduced until very recently when the large spiral-rich Virgo
II cloud has begun to fall into the cluster diluting the fraction of
early-type systems. The fact that Tully \& Shaya saw very few outwardly
moving galaxies outside the $6\degr$ VIC circle supported their
argument that most, or perhaps all, spirals and irregulars in Virgo,
mostly supplied by the Virgo II cloud, were recent arrivals.

All the findings of \citet{TS84} were based on a data set that
contained a limited number of galaxies with relatively uncertain
distance estimates. Now, with a much larger sample and more accurate
distances, we provide evidence that a substantial number of galaxies,
with a wide range of clustercentric distances, are expanding away from
Virgo. This suggests that these \hfill galaxies \hfill are \hfill probably \hfill reemerging \hfill after
\hfill infall. If \hfill our  
\begin{center}
\vspace{-3mm}
\includegraphics[width=0.95\columnwidth]{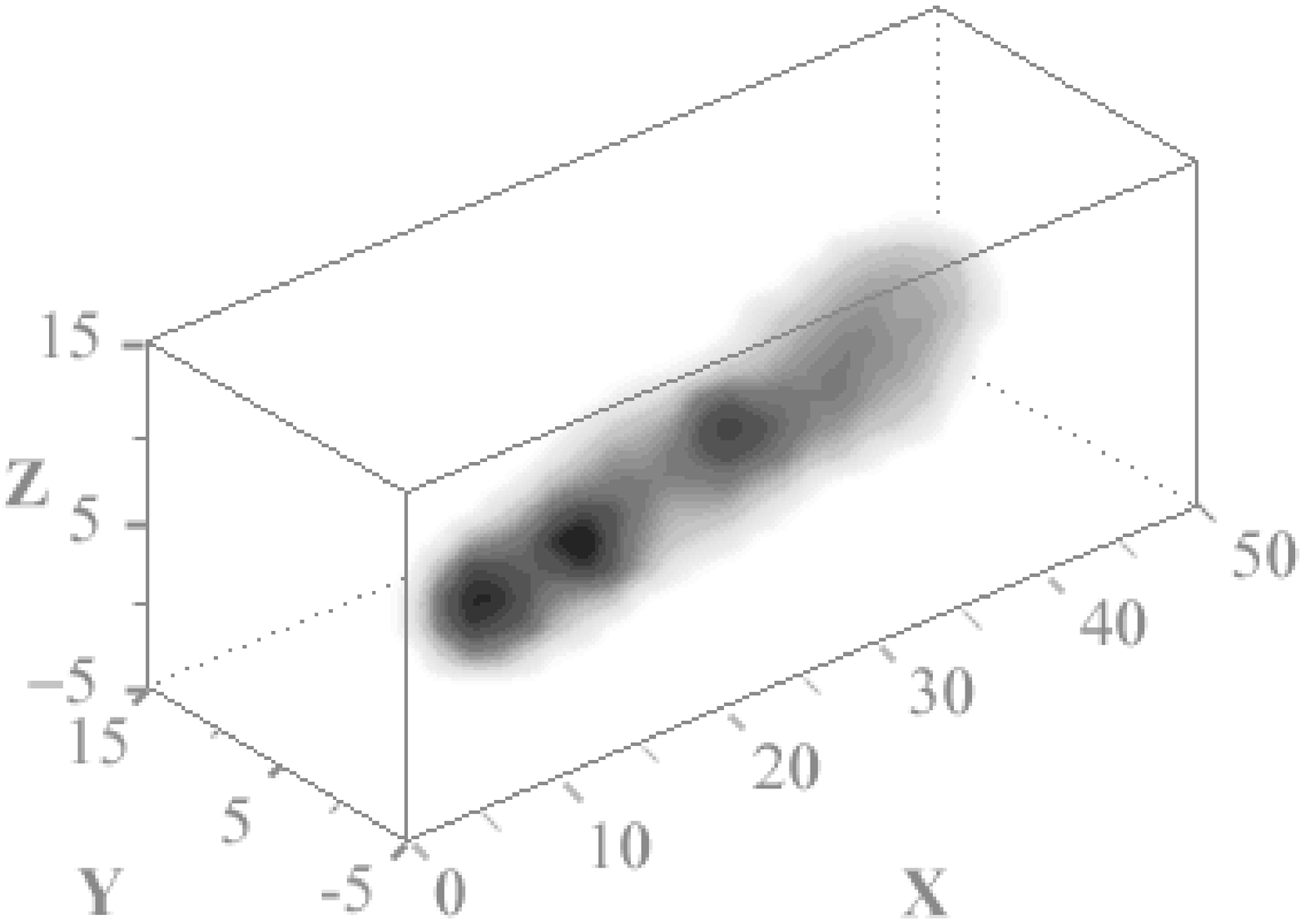}%
\makeatletter\def\@captype{figure}\makeatother
\figcaption{Voxel
projection of the 3D distribution of \hi\ deficiency in the VIC
region. The plot is in rectangular equatorial coordinates with
distances given in Mpc. The xy-plane corresponds to $\rm Decl.=0\degr$,
the x- and y-axis point to $\rm R.A.=12$ and 18 hr, respectively, and
the z-axis points to the north. The central dark spot is associated
with the cluster, with M87 being right at its center. The other two
enhancements are peripheral regions of \hi\ deficiency in the frontside
and backside of the VIC. Our position is at the origin of the
coordinate system.\label{voxel}}
\end{center}
interpretation of the situation is correct, then there is a more
continuous influx of galaxies into Virgo than previously anticipated,
so the hypothesis that the strong gaseous deficiencies currently seen
in some peripheral objects were originated in a previous infall episode
is worth exploring.

\section{The Main \hi\ Deficiency Enhancements of the Virgo 
Cluster}\label{structure3d}

A continuous representation in rectangular equatorial coordinates of
the spatial distribution of \hi\ deficiency in the VIC region is shown
in Figure~\ref{voxel}. This image is like a radiograph in which the
shade intensity informs on the average \hi\ deficiency of the galaxy
distribution observed from a given viewing angle. Three dark spots
indicating the accumulation of galaxies with a dearth of neutral
hydrogen are easily identified aligned along the line-of-sight
(LOS). In \citeauthor{Sol02} it is demonstrated that the central
enhancement of the \hi\ deficiency, which is essentially coincident
with the cluster core, arises from numerous gas-poor galaxies with LOS
distances ranging from \sm16 to 22 Mpc. Another region with important
\hi\ deficiency is associated with several nearby galaxies at LOS
distances \sm10--15 Mpc, preferentially located to the north of M87,
and moving away from the cluster with large relative
velocities. Finally, there is a tentative background group of galaxies
at LOS distances \sm25--30 Mpc, most with systemic velocities close to
the cluster mean, lying in the region dominated by the southern edge of
the classical M49 subcluster, and clouds W$^\prime$ and W.

While the frontside enhancement of the \hi\ deficiency is produced by
gas-poor galaxies that appear relatively clustered in 3D space simply
because they are nearby objects, the gas-deficient enhancement in the
background arises from a tight aggregation of galaxies segregated in
the four-dimensional position-radial velocity phase space. Up to 15 of
the galaxies listed in the data set in \citeauthor{Sol02} can be
identified as probable members of this group, since they all share
similar positions on the sky ($\mathrm{12^h15^m\le R.A.\le 12^h30^m}$\
and\ $\mathrm{+6\degr\le Decl.\le +10\degr}$). In addition, 12 of these
objects have LOS distances between 27 and 30 Mpc, and 8 of them (11 out
of the initial 15) have systemic velocities between \sm600--1300
\kms. Certainly, the lack of good resolution in the radial direction
prevents us for claiming that we have identified a true group on a
sufficiently safe basis. Yet, the fact that one third of its potential
members have gas deficiencies that deviate more than $2\sigma$ from
normalcy and that two of them have \hi\ masses less than 10\% of the
expectation values for their morphology ---characteristics that are
both typical of rich cluster interiors--- reinforces the impression
that it is not a fortuitous feature.

\section{Infall Model with Rebound}\label{model}

To model the virgocentric velocity field, we have implemented a simple
point mass model for the spherical collapse of a zero-pressure fluid
\citep[see, e.g.,][and references therein]{ET94}. However, the solution
of the problem is not limited to the classical deduction of the
velocity field of objects until a singularity first develops. The
calculations are extended beyond that time until galaxies recollapse
again by allowing that the spherical shells of matter reaching the
singularity pass through themselves and reemerge.

In the present paper, we take the, admittedly crude, point of view of
treating galaxies as test particles moving in a constant point-mass
potential well, which allows us to ignore the effects of shell crossing
on their first orbit. Shell crossing is most important inside the
previously relaxed cluster body, but galaxies on first infall spend
little time in this region because of their radial trajectories and
high pericentric velocities. Hence, the gravitational acceleration they
undergo during their first crossing of the cluster core is not expected
to differ substantially from that exerted if all the mass were
concentrated at the center.

Under the point mass approximation, the velocity relative to the center
of Virgo, $v$, for mass-less shells of radius $r$ in a universe with
null cosmological constant (a non-null cosmological constant has a
negligible contribution in the local universe) is given by the pair of
parametric equations

\begin{equation}\label{bound}
{\eta-\sin\eta\over{(1-\cos\eta)^{3/2}}}={\sqrt{GM_{\mathrm{VIC}}}t_0\over{r^{3/2}}}, \quad v={r\over{t_0}}{\sin\eta(\eta-\sin\eta)\over{(1-\cos\eta)^2}} 
\end{equation}
\\
for the bound shell case, and by

\begin{equation}\label{unbound}
{\sinh\eta-\eta\over{(\cosh\eta-1)^{3/2}}}={\sqrt{GM_{\mathrm{VIC}}}t_0\over{r^{3/2}}}, \quad v={r\over{t_0}}{\sinh\eta(\sinh\eta-\eta)\over{(\cosh\eta-1)^2}}
\end{equation}
\\ 
for the unbound case. In these expressions, $M_{\mathrm{VIC}}$ is the
\emph{effective} total mass of the VIC region (strictly speaking, the
point mass representative of the region under consideration around the
VIC center), $t_0$ is the current age of the universe, and $G$ is the
gravitational constant. In our double-infall model bound shells reach
maximum expansion for development angles $\eta=\pi$ and $3\pi$, whereas
full collapse is achieved when $\eta=2\pi$ and $4\pi$.

The predicted radial velocity, $V$, of a galaxy at a distance $R$ from
the Local Group (LG) and observed at an angular distance $\theta$ from
the Virgo center can be evaluated from

\begin{equation}\label{vobs}
V(R)=V_{\mathrm{VIC}}\pm v\sqrt{1-R_{\mathrm{VIC}}^2\sin^2\theta/r^2}\;,
\end{equation}
where $R_{\mathrm{VIC}}$ and $V_{\mathrm{VIC}}$ are, respectively, the
barycentric distance and velocity of the VIC in the LG reference
frame. The linear and angular distances of the galaxy are related
through the cosine law

\begin{equation}\label{coslaw}
d^2=1+D^2-2D\cos\theta\;,
\end{equation}
\\
where $d=r/R_{\mathrm{VIC}}$ and $D=R/R_{\mathrm{VIC}}$ are the linear
distances expressed in Virgo's distance units. The $(-)$-sign applies
for galaxies with $D<\cos\theta$, while the $(+)$-sign is for
$D\ge\cos\theta$.

\section{Model Predictions vs.\ Observations}\label{predictions}

As shown in the previous section, our simple dynamical model describing
the velocity field in the VIC region involves parameterization in terms
of $t_0$, $M_{\mathrm{VIC}}$, $R_{\mathrm{VIC}}$, and
$V_{\mathrm{VIC}}$ ---the systemic velocity can be replaced by the
cosmological velocity of Virgo for a given infall velocity of the
LG. In spite of the fact that only three of these parameters are
independent, it is advisable to work with no more than \emph{two} free
parameters, as the current velocity-distance data on the VIC are still
insufficient to constrain models with so much freedom. For this reason,
we allow the Virgo mass and distance to vary freely while keeping
$V_{\mathrm{VIC}}$ fixed to 980 \kms. We have, nonetheless, checked
that the range of values of $V_{\mathrm{VIC}}$ allowed by the
observations does not lead to significantly different results. The
viability of the solutions will be cross-checked by comparing the value
of $t_0$ predicted by the best fits to the observations with the
cosmological age of $13\pm 1$ Gyr derived for the
$\Omega_{\mathrm{m}}=0.3$, $\Omega_\Lambda=0.7$ CDM model in the
$H_0$-Key Project by \citet{Fre01} and the more precise $t_0=13.6\pm
0.2$ Gyr recently inferred by \citet{Sie02} from CMB observations.

The systemic velocity-distance diagram for the VIC region plotted in
Figure~\ref{hubdiam}a shows the basic expected features: an initial
steeply rising velocity-distance relation at the cluster front, a
central broad region with the maximum observed velocity amplitudes, and
a final ascending part of the relation, expected to approach the local
Hubble law asymptotically. Ultimately, all the available good-quality
observations should be considered to define the constraints of
dynamical models like ours. Unfortunately, as illustrated in
Figure~\ref{hubdiam}b, even after we have extended the predictions to
the first orbit following rebound, the theoretical pattern cannot
explain the motions of various systems within the post-rebound
expansion regime. This fact and the difficulties inherent to the
modeling of the motions in the innermost cluster region ---where
multiple rebounds are expected to occur--- prompted us to restrict the
acceptable range of models by putting all the weight of the fits on the
envelope to the streaming velocities \emph{within the asymptotic
regime} (cf.\ Fig.~\ref{hubdiam}b). This is indeed the fitting
procedure adopted by first-infall models and has the advantages of: (1)
relying on the subset of observations that offer the highest
constraining power; (2) being unaffected by shell crossing; and (3)
being barely sensitive to the angular distance from the VIC center.

The three solid curves in Figure~\ref{hubdiam}a demonstrate the locus
of the optimum model for lines-of-sight corresponding to angular
separations of 4$\degr$, 6$\degr$, and 8$\degr$. We find it compelling
that, in spite of the fact that the minimization relies on a fraction
of the data, our best solution explains a large number of the galaxy
motions around the VIC region. The good accordance between the
observations and the prediction of the double-infall model is
reinforced by the remarkable symmetry of the motions with respect to
the local Hubble flow defined by the imaginary straight
\begin{center}
\includegraphics[bb=100 104 674 678, width=1.35\columnwidth]{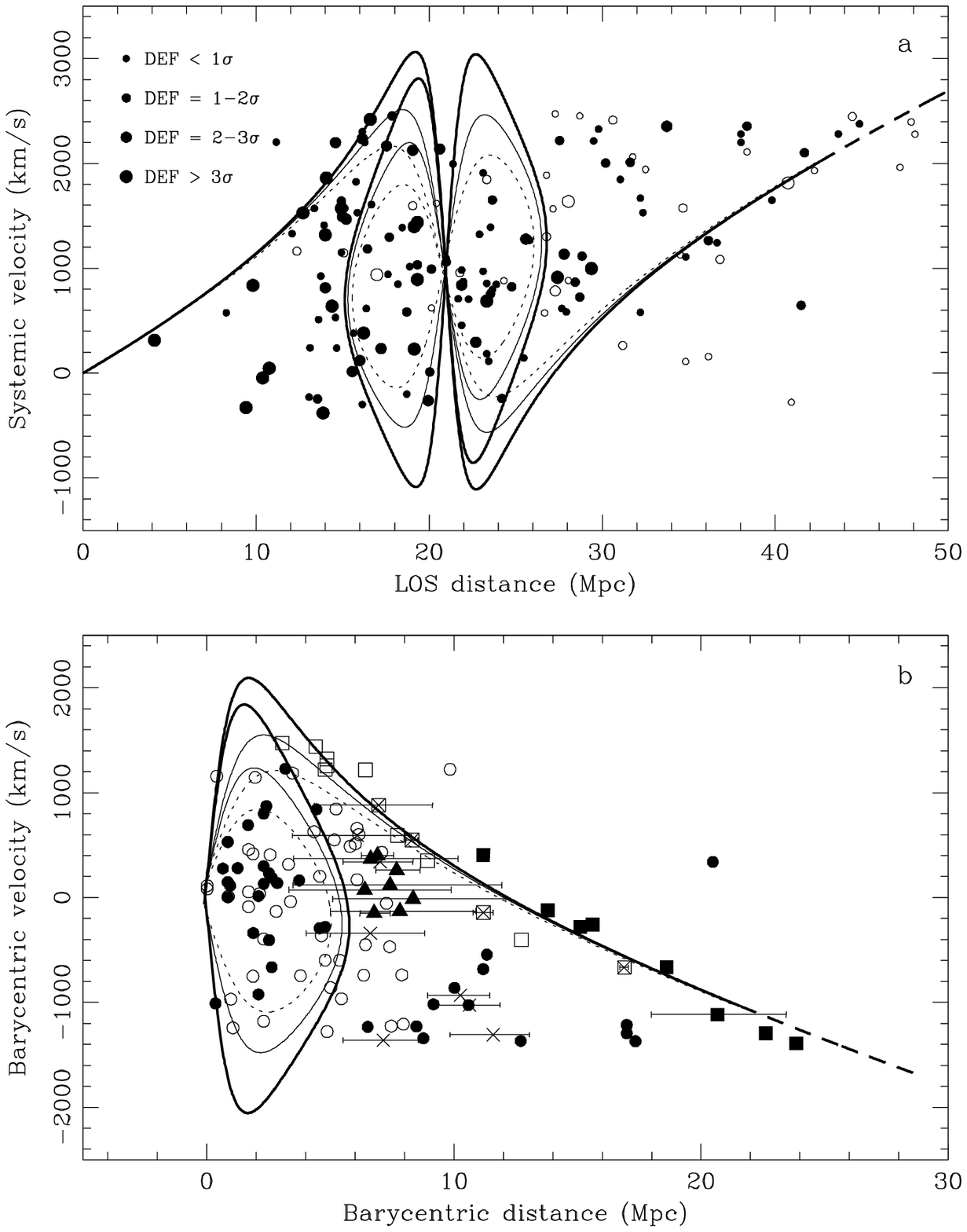}%
\makeatletter\def\@captype{figure}\makeatother
\vspace{-12mm}
\figcaption{(a) Systemic velocity vs.\ LOS distance
for Virgo spirals. The curves demonstrate the predicted velocities for
$\theta=4\degr$ (thick solid), $\theta=6\degr$ (thin solid), and
$\theta=8\degr$ (dotted). The dashed portion at large Virgocentric
distances is for unbound shells. Galaxies with uncertain distances
(open symbols) have been excluded from the fit. Parameter \df\ measures
\hi\ deficiency in units of the mean standard deviation for field
objects ($=0.24$). (b) Barycentric velocity vs.\ barycentric distance
for spirals with reliable LOS distances. Open and filled symbols are,
respectively, for galaxies closer and more distant than the
barycenter. Squares identify the galaxies used to define the two
asymptotic branches, triangles are for members of the \hi-deficient
background group, and crosses for galaxies with $\df\geq 3\sigma$ and
LOS distance $\leq 15$ Mpc. Error bars in distance are shown for the
last two subsets of galaxies and for one of the galaxies within the
asymptotic regime. They are representative of the uncertainty on the
distances of all the available good-quality observations. Errors in the
velocities are negligible in comparison. The curves are the same
solutions depicted in (a).\label{hubdiam}}
\end{center}
\vspace{-3mm}
line passing through the position of the LG and that of 
the predicted VIC barycenter (cf.\ Fig.~\ref{hubdiam}b). Moreover, the
error contours in the $(M_{\mathrm{VIC}},R_{\mathrm{VIC}})$ plane drawn
in Figure~\ref{contours} show that our predictions are in reasonable
agreement with previous estimates of the values of these
parameters. Thus, while mass calculations of the central cluster region
based on the X-ray emission or the virial theorem amount to about
several times $10^{14}$ M$_\odot$, our best estimate
$M_{\mathrm{VIC}}=2.8\times 10^{15}$ M$_\odot$ approaches the values
$\lesssim 2\times 10^{15}$ M$_\odot$ inferred from modelings of the
velocity field of the Local Supercluster \citep{TS84,Fou01}. The
acceptable barycentric distances are also well within the very poorly
constrained range (\sm16--24 Mpc) of VIC distances reported in the
literature, although our most likely value $R_{\mathrm{VIC}}=21.0$ Mpc
advocates a large-distance scale which, for a cosmological VIC velocity
of 1200 \kms, brings the \emph{local} value of the Hubble constant to
about 60 km/s/Mpc. Furthermore, our best solution leads to $t_0=13.5$
Gyr, in excellent agreement with the expansion ages advocated in
\citet{Fre01} and \citet{Sie02}. Finally, note from eq.~(\ref{vobs})
that the constraint provided by the motion of the LG with respect to
the VIC (i.e., a null systemic velocity at our position) 
is automatically satisfied by the models.

\begin{center}
\includegraphics[width=0.9\columnwidth]{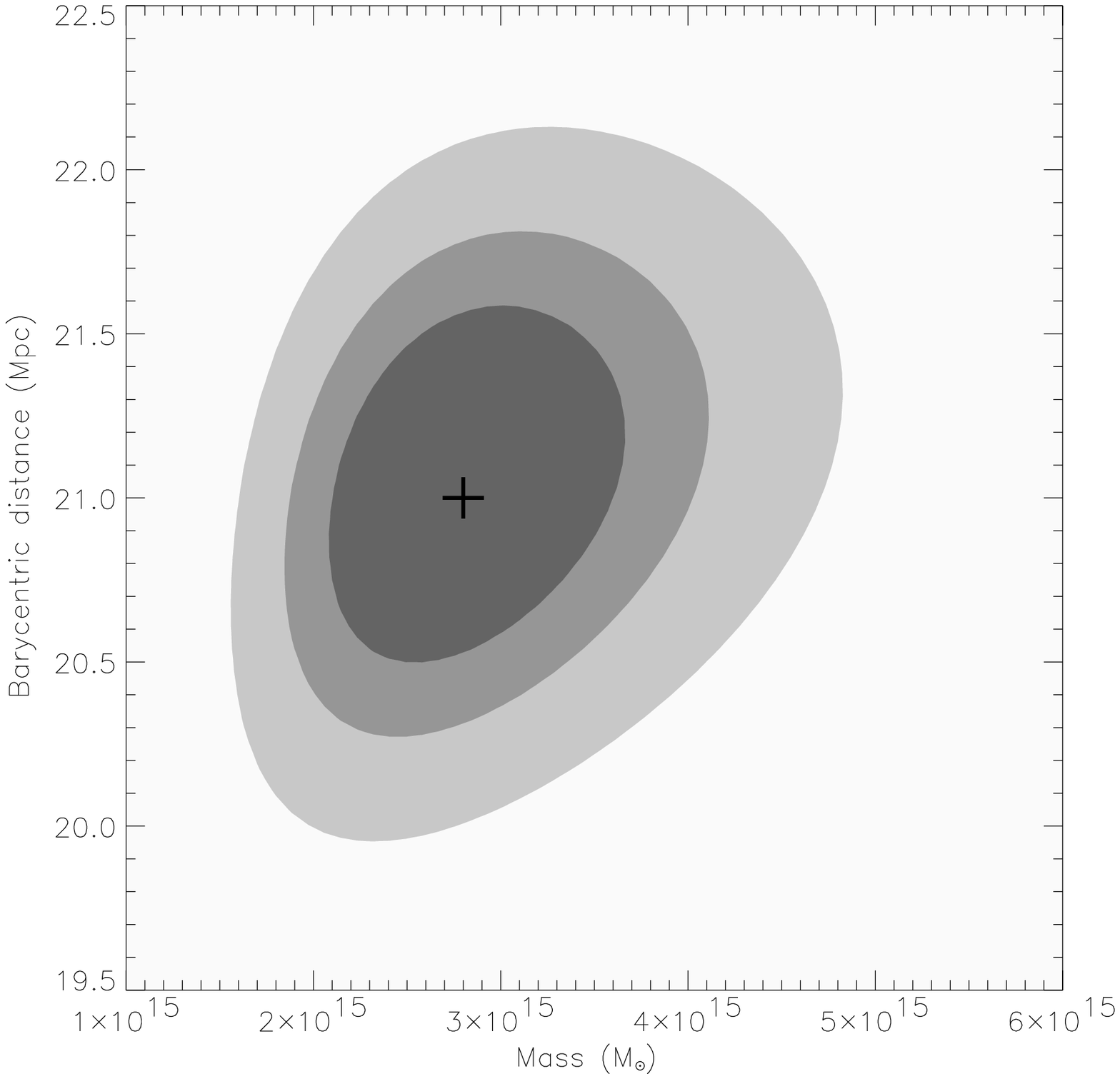}%
\makeatletter\def\@captype{figure}\makeatother 
\vspace{2mm}
\figcaption{1, 2, and $3\sigma$ confidence limits on 
$M_{\mathrm{VIC}}$ and $R_{\mathrm{VIC}}$ obtained from our dynamical
model of the VIC region. The cross indicates the best estimates of
these parameters.\label{contours}}
\end{center}

\section{Implications on the Origin of the Gas Deficiency}

One of the most striking predictions of our model, illustrated in
Figure~\ref{hubdiam}b, is that various \hi-deficient galaxies in the
frontside and backside of Virgo can be accommodated in the same way on
an orbit following rebound and on a first infall trajectory. We
conclude that these galaxies might not be recent arrivals but have
already plunged into the VIC center in the past. Especially interesting
is the case of the \hi-deficient background group observed at $R\sim
28$ Mpc and $V\sim 1000$ \kms, whose near turnaround position in the
Hubble diagram and the assumed standard harmonic oscillation movement
of the galaxies around the cluster center, together with a cosmic age
presumably close to 13.5 Gyr, indicate that it might have passed
through the Virgo core about 4.5 Gyr ago. The fit to the observed
velocity field also indicates that a significant number of galaxies in
the VIC region might be currently falling towards the cluster from the
frontside ---perhaps not for the first time either---, so it does not
substantiate former claims of a paucity of objects of this sort
\citep{Fou01}.

We do not claim, however, that this is the only possible solution
satisfying the observations. For instance, an alternative model based
on the assumption that the gas-deficient group in the background is
also on first infall gives similarly reasonable best values of the
assumed parameters: $M_{\mathrm{VIC}}=1.8\times 10^{15}$ M$_\odot$ and
$R_{\mathrm{VIC}}=20.2$ Mpc, leading to $t_0=14.0$ Gyr. Nor are we
endorsing the argument that \emph{all} \hi-deficient galaxies in the
VIC region have passed through the cluster core. It is clear from
Figures~\ref{hubdiam}a,b that the positions of some gas-poor galaxies,
especially in the front cluster side, are best explained according to
our model if they are on first infall. In any event, our dynamical
modeling of the VIC region has shown that characteristics such as a
large Virgocentric distance or a near turnaround position are not by
themselves conclusive indications of a recent arrival. The substantial
\hi\ deficiency of the background subclump found in \citeauthor{Sol02}
may well have originated in an earlier passage of this entity through
the Virgo core.

Even if our suggestion that the \hi-deficient group on the backside of
the VIC might not be a recent arrival is finally proven well-founded,
it is still necessary to find an explanation for the apparently long
time (\sm 4.5 Gyr) that these galaxies have maintained a substantial
dearth of gas without noticeable consequences on their morphologies: 9
of its 15 probable members are late-type spirals, whereas the 5 with
the largest gaseous deficiencies have types Sb or later. (Nor do the
magnitudes of these latter galaxies show any evidence of a substantial
dimming.) Yet, the details and chronology of the evolution of galactic
properties triggered by the sweeping of the atomic hydrogen, as well as
its repercussions on the star formation rate, are still poorly
understood. For instance, estimates for gas consumption time scales in
the absence of gas replenishment from a sample of 36 spiral galaxies of
various morphologies by \citet*{LTC80} produced values ranging from 0.9
to 15 Gyr, with a median of 3.9 Gyr. These authors also discussed the
color evolution of disk galaxies whose star formation has been
truncated at various past times and concluded that substantial
reddening in the latest spiral types requires that most of their star
formation ceased about 5 Gyr ago. On the other hand, recent
observational studies of the galaxy populations in clusters at
different redshifts \citep[e.g.,][]{Fas00} indicate that while the
quenching of star formation induced by the removal of the gas appears
to be rapid (\sm1 Gyr), the morphological evolution of disk galaxies
takes several billion years. Perhaps evolution is still slower for
objects, such as the members of the gas-deficient cloud detected in the
background of Virgo, that spent most of their time out of the
aggressive cluster environment.

\begin{acknowledgements}
We acknowledge Brent Tully for stimulating conversation and comments,
and Montserrat L\'opez for assistance with Figure~\ref{voxel}. This
work was supported by the Direcci\'on General de Investigaci\'on
Cient\'{\i}fica y T\'ecnica, under contract AYA2000--0951. TS
acknowledges support from a fellowship of the Ministerio de
Educaci\'on, Cultura y Deporte of Spain.
\end{acknowledgements}

\singlespace


\begin{thebibliography}{} 
\bibitem[Bravo-Alfaro et al.(2000)]{Bra00} Bravo-Alfaro, H., Cayatte, V., van Gorkom, J.~H., \& Balkowski, C.\ 2000, \aj, 119, 580 
\bibitem[Cayatte et al.(1994)]{Cay94} Cayatte, V., Kotanyi, C., Balkowski, C., \& van Gorkom, J.~H.\ 1994, \aj, 107, 1003
\bibitem[Dickey \& Gavazzi(1991)]{DG91} Dickey, J.~M., \& Gavazzi, G.\ 1991, \apj, 373, 347
\bibitem[Ekholm \& Teerikorpi(1994)]{ET94} Ekholm, T., \& Teerikorpi, P.\ 1994, \aap, 284, 369
\bibitem[Fasano et al.(2000)]{Fas00} Fasano, G., Poggianti, B.~M., Couch, W.~J., Bettoni, D., Kj\ae rgaard, \& P., Moles, M.\ 2000, \apj, 542, 673
\bibitem[Fouqu\'e et al.(2001)]{Fou01} Fouqu\'e, P., Solanes, J.~M., Sanchis, T., \& Balkowski, C.\ 2001, \aap, 375, 770
\bibitem[Freedman et al.(2001)]{Fre01} Freedman, W.~L., et al.\ 2001, \apj,  553, 47 
\bibitem[Gavazzi \& Jaffe(1987)]{GJ87} Gavazzi, G., \& Jaffe, W. 1987, \aap, 186, L1
\bibitem[Giovanelli \& Haynes(1985)]{GH85} Giovanelli, R., \& Haynes, M.~P.\ 1985, \apj, 292, 404 
\bibitem[Larson et al.(1980)Larson, Tinsley, \& Caldwell]{LTC80} Larson, R.~B., Tinsley, B.~M., \& Caldwell, C.~N.\ 1980, \apj, 237, 692
\bibitem[Quilis et al.(2000)Quilis, Moore, \& Bower]{QMB00} Quilis, V., Moore, B., \& Bower, R.\ 2000, \sci, 288, 1617
\bibitem[Sievers et al.(2002)]{Sie02} Sievers, J.~L., et al.\ 2002, astro-ph/0205387
\bibitem[Solanes et al.(2001)]{Sol01} Solanes, J.~M., Manrique, A., Gonz\'alez-Casado, G., Garc\'\i a-G\'omez, C., Giovanelli, R., \& Haynes, M.~P.\ 2001, \apj, 548, 97 
\bibitem[Paper~I(2002)Solanes et al.]{Sol02} Solanes, J.~M., Sanchis, T., Salvador-Sol\'e, E., Giovanelli, R., \& Haynes, M.~P.\ 2002, \aj, submitted (Paper~I)
\bibitem[Stevens et al.(1999)Stevens, Acreman, \& Ponman]{SAP99} Stevens, I.~R., Acreman, D.~M., \& Ponman, T.~J. 1999, \mnras, 310, 663
\bibitem[Tully \& Shaya(1984)]{TS84} Tully, R.~B., \& Shaya, E.~J. 1984, \apj, 281, 31 
\bibitem[van Dokkum et al.(1999)]{vDok99} van Dokkum, P.~G., Franx, M., Fabricant, D., Kelson, D.~D., \& Illingworth, G.~D. 1999, \apj, 520, L95 
\bibitem[Vollmer et al.(2001)]{Vol01} Vollmer, B., Cayatte, V., Balkowski, C., \& Duschl, W.~J.\ 2001, \apj, 561, 708
\end{thebibliography}
\end{document}